\documentclass[10pt,letterpaper]{article}
\usepackage[top=0.85in,left=2.75in,footskip=0.75in]{geometry}

\usepackage{amsmath,amssymb}

\usepackage{changepage}

\usepackage[utf8]{inputenc}

\usepackage{textcomp,marvosym}

\usepackage{nameref,hyperref}

\usepackage[right]{lineno}

\usepackage{microtype}
\DisableLigatures[f]{encoding = *, family = * }

\usepackage[table]{xcolor}

\usepackage{array}

\newcolumntype{+}{!{\vrule width 2pt}}

\newlength\savedwidth

\newcommand\thickhline{\noalign{\global\savedwidth\arrayrulewidth\global\arrayrulewidth 2pt}%
\hline
\noalign{\global\arrayrulewidth\savedwidth}}

\raggedright
\usepackage[aboveskip=1pt,labelfont=bf,labelsep=period,justification=raggedright,singlelinecheck=off]{caption}

\usepackage{lastpage,fancyhdr,graphicx}
\usepackage{epstopdf}
\fancyheadoffset[L]{2.25in}
\fancyfootoffset[L]{2.25in}
\begin{document}
\vspace*{0.2in}

\begin{flushleft}
{\Large
\textbf\newline{Industrial Topics in Urban Labor System} 
}
\newline
\\
Jaehyuk Park \textsuperscript{1,2,3},
Morgan R. Frank \textsuperscript{4,5,6},
Lijun Sun \textsuperscript{7},
Hyejin Youn \textsuperscript{1,2,8*},
\\
\bigskip
\textbf{1} Kellogg School of Management, Northwestern University, Evanston, IL, USA
\\
\textbf{2} Northwestern Institute on Complex Systems, Evanston, IL, USA
\\
\textbf{3} Luddy School of Informatics, Computing, and Engineering, Indiana University, Bloomington, IN, USA
\\
\textbf{4} Department of Informatics and Networked Systems, School of Computing and Information, University of Pittsburgh, Pittsburge, PA, USA
\\
\textbf{5} Connection Science, Institute for Data, Systems, and Society, Massachusetts Institute of Technology, Cambridge, MA, USA
\\
\textbf{6} Digital Economy Lab, Institute for Human-Centered AI, Stanford University, Stanford, CA, USA
\\
\textbf{7} Department of Civil Engineering, McGill University, Montreal, QC, Canada 
\\
\textbf{8} Department of Civil and Environmental Engineering at Northwestern University, Evanston, IL 60208, USA.

\bigskip

%
%





* hyejin.youn@kellogg.northwestern.edu

\end{flushleft}
\section*{Abstract}

Categorization is an essential component for us
to understand the world for ourselves and to communicate it collectively.
It is therefore important to recognize that classification systems is not necessarily static, especially for economic systems, and even more so in urban areas where most innovation takes place and is implemented. Out-of-date classification systems would potentially limit further understanding of the current economy because things constantly change.  Here, we develop an occupation-based classification system for the US labor economy, called industrial topics, that satisfy adaptability and representability. 
By leveraging the distributions of occupations 
across the US urban areas, we identify industrial topics --- 
clusters of occupations based on their co-existence pattern. 
Industrial topics indicate the mechanisms under the systematic 
allocation of different occupations. Considering the densely connected 
occupations as an industrial topic, our approach characterizes regional economies by their topical composition.
Unlike the existing survey-based top-down approach, 
our method provides timely information about 
the underlying structure of the regional economy, which is critical for policymakers and business leaders 
especially in our fast-changing economy. 


\section*{Introduction}
From ancient civilization, human beings has long been classifying
things around themselves: the edible or poisonous, 
or the dangerous or useful. Categorization does not only
help our brain to learn and understand the outside world individually, 
but also equips us with efficient ways to share and maintain individually acquired knowledge
with the community. The latter is an essential component for the advancement of science and technology, playing, therefore, a crucial role, 
if not indispensable, for a society to survive and thrive \cite{sutherland2015memory, liberman2017origins}.
The better we understand a system, the more sophisticated its classification becomes, thus providing a better toolkit for further understanding. 
In biology, for example, taxonomy --- the system of categorizing and describing species --- does not just provide a list of species as a result of ``the pioneering exploration of life on a little known planet'' \cite{wilson2004taxonomy}, but also helps to find 
testable scientific predictions \cite{agnarsson2007taxonomy}.
Had not been Aristotle's {\it scala naturae} (great chain of things) and {\it Linnaean taxonomi}, for example, it would have taken much longer time for Darwin and Wallace to develop the evolutionary theory from numerous empirical observations \cite{voultsiadou2017aristotle}. 
It is therefore reasonable to say that our scientific progress has been
built upon classification systems. 


There are two characteristics that a \emph{good} classification system must satisfy. 
First, the classification system has to be adaptive to changes in the system. 
Indeed, many socioeconomic systems, including urban systems, are never static, but constantly creating novel services and products to increase diversity and productivity. 
When newly created services and products are truly innovative, we are often unable to place them in an existing classification scheme, and hence need either to create a new \emph{ad hoc} category, or to revise the existing classifications to find their home  \cite{youn2016, bettencourt2014}. 
Second, a meaningful classification system must capture the variation at the most meaningful resolution for the required unit of analysis, and then arrange the variation to fulfil functional needs \cite{youn2016}.
Take a furniture store, for example. When we walk into a store, furniture is primarily laid out by their functional purposes, such as chairs, tables, and beds, and then further down to by their styles and material types, in ways to fulfill our primary needs most efficiently. 

Urban economy is best known for its fast pace of bringing about 
novel products and economic services \cite{bettencourt2007, frank2018}. 
When created products and services are truly innovative, the contemporary classification is most likely to fail to fully appreciate novel features \cite{youn2016}. 
For example, the U.S, North American Industry Classification System (NAICS) 
is revised every five years \footnote{https://www.census.gov/eos/www/naics/index.html}, which is is not necessarily a timely revision. 
In addition, the revision needs to respect past classification decisions, for which it often misses the most valuable companies in the world as Libert et al. states in the following~\cite{libert2016we}: 


\begin{quote}
Consider the five most valuable companies in the world, according to S\&P: 
Apple, Alphabet (Google), Amazon, Microsoft, and Facebook, which we call the 
Fab Five. Despite the fact that these companies make money in different ways — 
Apple makes most of its money on hardware, Microsoft on software, and Facebook 
and Google from advertising — they do share a lot of similarities. 
But Information Technology doesn't seem like the right category to group them 
into. The fifth member of the group, Amazon, is officially a Consumer 
Discretionary firm. There’s no denying that it is a retailer, but it also has a 
digital platform rather than physical stores, and nearly 50\% of the units sold 
through its website are sold by third-party sellers. With that in mind, 
Amazon seems closer in DNA to Facebook than Walmart. Like Facebook, it created 
an open platform that anyone, anywhere in the world, can use.
\end{quote}

The current industrial classification systems use physical 
firms or tradable goods as the fundamental unit 
while the global economy is increasingly driven by human capital with knowledge and skill, 
as a result of top-down decisions by the Economic Classification Policy Committee (ECPC)  that creates categories 
to best serve tradable goods and businesses. 
For instance, the most widely used industrial classification, NAICS, is constructed
based on the product of business establishments \cite{ambler_introducing_1998}.    
However, as the global economy is increasingly relying on knowledge and services, the 
importance of intangible inputs and outputs --- such as the types of skills, 
occupations, and produced services --- has increased in characterizing and 
analyzing the economic activity of a region~\cite{frank2019toward}. In sum, the lack of 
occupation-based classification system limits the analysis of the economic structure 
of a nation or a region in terms of its distribution of human capital. 

We apply a machine learning algorithm to identify the latent topic 
from the distributions of occupations across the US urban areas.
The topic modeling algorithms has been widely to find a latent structure in large-scale documents,
including recommendation system and computational social science.
We identify industrial topics as latent clusters of occupations that
characterize and relate the regional economies.  
The comparison between big and small cities shows the 
structural difference in the urban economy and the prevalence map of each 
topic reveals the distributional pattern of each industrial topic. 
Our method has a significant benefit when it comes to the temporal evolution of urban economy \cite{Hong2020}. 
Furthermore, the change of our industrial topics over time provides us with the structural 
dynamics of the national economy through the lens of labor. 

The computational approach in the characterization of a regional economy has 
significant academic contribution in both theory and methodology. 
Considering an industrial topic as a latent group of occupations, 
the topic offers a new holistic lens that links occupation and industry.
We expect that the new lens can provide a new systematic approach to solve essential 
questions in the intersection of labor economics and economic structure, 
such as the local employment multipliers \cite{thompson1959investigation, 
weiss1970estimation, moretti2010local, van2017local}. 
Furthermore, the computational approach can provide a more economic and 
timely information about the underlying structure of the regional economy, 
compared to the existing survey-based top-down approach, which becomes more 
important for policymakers and business leaders in the fast-changing economy.

\section*{Results}
Our approach begins with an analogy between a document and a regional economy. 
A document consists of the words inside. Accordingly, if we can find a 
set of words that are likely to locate in the same document, we can 
estimate the topic of the documents as the composition of the words. Moreover, 
by extracting multiple word sets from documents, we can see 
what topics exist across the documents, as well as what topics are 
over-represented in a certain document. Under this logic, computer scientists 
and linguisticians have worked on developing computational algorithms, 
topic modeling, to extract the topic —-- as a composition of words —-- 
from a large-scale set of documents.

Similarly, in a regional economy, different occupations are distributed 
unevenly, based on the strength of the regional economy.
Hence, if we can find a group of occupations that frequently appear together 
in different regions, we can also call it an industrial topic. Furthermore, 
similar to the aforementioned document-topic relationship, the list of topics 
across all regions in a country can reveal the list of industrial topics that 
are regionally specialized, while we can also characterize a certain regional 
economy as a combination of those industrial topics.

Based on this correspondence, we apply 
\textit{Nonnegative Matrix Factorization (NMF)} to 
the occupation distribution records of the metropolitan areas in the U.S., 
to extract an industrial topic as a group of occupations. Our approach 
using topic modeling method allows us to redefine the concept of the industry as 
the group of occupations, and to characterize the regional economic 
structures more organically and efficiently. In particular, since the topic 
modeling algorithm infers a latent topic based on locally allocated occupations, 
we focus on the extraction of the traded industries, defined as the industries 
concentrating in particular regions but sell products or services across 
regions and countries \cite{porter2003, delgado2014categorization}.  
A recent study finds that traded industries accounted for 36.0\% of total 
U.S. employment, 50.5\% of payroll, and about 91.2\% of patenting activity in 
2009 \cite{delgado2014categorization}.

\subsection*{Over-represented occupations in urban area}

\begin{table}[!t]
\begin{adjustwidth}{-1.25in}{0in} 
\centering
\small
\caption{
{\bf Four regions and their over-represented occupations}}
\begin{tabular}{|c|l|}
\hline
{\bf Region} & {\bf Top Five Occupations (in order)}\\ \thickhline
    & Flight Attendants \\
    & Actors \\
    Los Angeles-Long Beach-Anaheim, CA & Film and Video Editors \\
    & Media and Communication Equipment Workers, All Other \\
    & Media and Communication Workers, All Other \\\hline
    & Computer Hardware Engineers \\
    & Software Developers, Systems Software \\
    San Jose-Sunnyvale-Santa Clara, CA & Semiconductor Processors \\
    & Software Developers, Applications \\
    & Electronics Engineers, Except Computer \\\hline
    & Gaming Cage Workers \\
    & Gaming Dealers \\
    Las Vegas-Henderson-Paradise, NV & Gaming Supervisors \\
    & Gaming Service Workers, All Other \\
    & First-Line Supervisors of Gaming Workers \\\hline
    & Economists \\
    & Political Scientists \\
    Washington-Arlington-Alexandria, DC-VA-MD-WV & Legal Support Workers, All Other \\
    & Social Scientists and Related Workers, All Other \\
    & Flight Attendants \\\hline
\end{tabular}
\begin{flushleft} The occupations having the highest TF-IDF scores of four urban areas
 are matched with prevalent industries of the regions.
\end{flushleft}
\label{table1}
\label{tab:tf_idf}
\end{adjustwidth}
\end{table}

All regional economies include local occupations that provide goods and services 
primarily to the local market, similar to the functional words that are 
frequently used across all different documents (e.g., articles, pronouns, and 
conjunctions). Restaurant servers and retail managers can be examples of 
local occupations whose employment size are known as roughly proportional to 
the regional population \cite{porter2003}.

To mitigate the effect of local occupations and to normalize the size effect 
across regions, we re-weigh the employment size of each occupation in a region 
using Term Frequency–Inverse Document Frequency (TF-IDF). TF-IDF is a quantity
assigned to each pair of a word and a document, which captures 
how important the word is to 
the document in a collection of documents \cite{rajaraman2011mining}. 
In our application to the regional occupational distribution, the TF-IDF value 
increases proportionally to the number of times an occupation appears in the 
region. It is offset by the number of regions in all MSAs that have the 
occupation, which helps to adjust for the fact that local occupations appear 
more frequently in general (See Materials and methods for detail). 

Our result shows that the TF-IDF scores successfully extract the over-represented 
occupations in a region. To validate how effective TF-IDF score is to extract 
the representative occupations in a region, we check the top five occupations 
in TF-IDF score for the four regions, whose specialized industries are 
well-known: Los Angeles-Long Beach-Anaheim, CA, San Jose-Sunnyvale-Santa Clara, 
CA, Las Vegas-Henderson-Paradise, NV, and Washington-Arlington-Alexandria, 
which is located across DC, VA, MD, and WV. 
As shown in Table~\ref{tab:tf_idf}, the over-represented occupations 
in the four regions are well-matched with the corresponding industries for which
the regions are known. For instance, the occupations related to entertainment 
industry, such as film and video editors and media \& communication workers, 
have high TF-IDF scores in Los Angeles-Long Beach-Anaheim, CA --- 
where Hollywood is, while the occupations in computer hardware and software 
industries have high scores in San Jose-Sunnyvale-Santa Clara, CA --- 
where Silicon Valley locates.
Similarly, the occupations with high TF-IDF scores in Las Vegas and 
Washington DC areas are also well corresponded to the specialized industries 
of the regions. 

\subsection*{Industrial topics}

\begin{table}[!t]
\begin{adjustwidth}{-2.25in}{0in} 
\centering
\caption{
{\bf First five industrial topics and their prevalent occupations}}
\begin{tabular}{|c|l|}
\hline
{\bf Topic} & {\bf Prevalent Occupations}\\ \thickhline
     1 &  Textile Winding, Twisting, And Drawing Out Machine Setters, Operators, And Tenders; Textile Knitting\\
      & And Weaving Machine Setters, Operators, And Tenders; Extruding And Forming Machine Setters, Operators,\\
        & And Tenders, Synthetic And Glass Fibers; Textile Bleaching And Dyeing Machine Operators And Tenders;\\
        & Textile Cutting Machine Setters, Operators, And Tenders; Textile, Apparel, And Furnishings Workers, \\
        & All Other; Cutters And Trimmers, Hand; Sewing Machine Operators; Fabric Menders, Except Garment; \\
        & Chemical Equipment Operators And Tenders; Machine Feeders And Offbearers; Coating, Painting, And \\
        & Spraying Machine Setters, Operators, And Tenders \\\hline
     2 &  Farmworkers And Laborers, Crop, Nursery, And Greenhouse; First-Line Supervisors Of Farming, Fishing,\\
     & And Forestry Workers; Agricultural Equipment Operators; Graders And Sorters, Agricultural Products;\\
        & Substitute Teachers; Farmers, Ranchers, And Other Agricultural Managers; Separating, Filtering, \\
        & Clarifying, Precipitating, And Still Machine Setters, Operators, And Tenders; Psychiatric Technicians;\\
        & Farm Equipment Mechanics And Service Technicians; Demonstrators And Product Promoters; Detectives\\
        & And Criminal Investigators; Plasterers And Stucco Masons; Power Plant Operators \\\hline
     3 &  Service Unit Operators, Oil, Gas, And Mining; Roustabouts, Oil And Gas; Petroleum Engineers; \\
       & Derrick Operators, Oil And Gas; Wellhead Pumpers; Rotary Drill Operators, Oil And Gas; \\
        & Helpers--Extraction Workers; Petroleum Pump System Operators, Refinery Operators, And Gaugers;\\
        & Geological And Petroleum Technicians; Pump Operators, Except Wellhead Pumpers; Geoscientists, \\
        & Except Hydrologists And Geographers; Control And Valve Installers And Repairers, Except Mechanical \\
        & Door; Riggers; Insulation Workers, Mechanical; Gas Plant Operators;\\\hline
     4 &  Gaming Dealers; Gaming Supervisors; Gaming Change Persons And Booth Cashiers; First-Line Supervisors\\
      & Of Gaming Workers; Entertainers And Performers, Sports And Related Workers, All Other; Gaming Cage\\
        & Workers; Gaming Service Workers, All Other; Gaming Surveillance Officers And Gaming Investigators;\\
        & Gaming Managers; Baggage Porters And Bellhops; Parking Lot Attendants; Personal Care And Service \\
        & Workers, All Other; Audio And Video Equipment Technicians; Food Preparation And Serving Related \\
        & Workers, All Other; Slot Supervisors \\\hline
     5 &  Captains, Mates, And Pilots Of Water Vessels; Sailors And Marine Oilers; Riggers; Ship Engineers;\\
     & Service Unit Operators, Oil, Gas, And Mining; Petroleum Pump System Operators, Refinery Operators, \\
        & And Gaugers; Roustabouts, Oil And Gas; Bridge And Lock Tenders; Crane And Tower Operators; \\
        & Plant And System Operators, All Other; Commercial Pilots; Marine Engineers And Naval Architects; \\
        & Rotary Drill Operators, Oil And Gas; Pump Operators, Except Wellhead Pumpers; Structural Metal\\
        & Fabricators And Fitters \\\hline
\end{tabular}
\begin{flushleft} The first five industrial topics decided by our NMF algorithm, and the ten most prevalent occupations for each topics. The rest of the topics (From Topic 6 to 10) and their corresponding prevalent occupations are presented in Supporting information.
\end{flushleft}
\label{tab:nmf_ex}
\end{adjustwidth}
\end{table}

Based on the TF-IDF score for each occupation across all metropolitan areas in 
the US, we extract the latent industrial topics using 
Nonnegative Matrix Factorization (NMF), one of the most widely used topic 
modeling algorithms. 
Here, according to our analogy between a regional economy and a document, 
we apply NMF to extract the industrial topic defined as a group of occupations 
that are likely to be located together (See Materials and methods for more 
detail). 

Our results show that the topic modeling algorithm can successfully extract 
the industrial topics —-- especially those of the traded industries --— and 
their representative occupations. As presented in 
Table~\ref{tab:nmf_ex}, which lists the first five industrial topics with their 
ten most representative occupations, each topic consists of a group of closely 
related occupations to a specific industry. For instance, the representative 
occupations of Topic 1 are the ones in the textile industry, such as ``Textile 
Winding, Twisting, and Drawing Out Machine Setters, Operators, and Tenders'', 
``Textile Knitting and Weaving Machine Setters, Operators, and Tenders'',  
and ``Textile Bleaching and Dyeing Machine Operators and Tenders''. Similarly, 
we can easily match the other topics in Table~\ref{tab:nmf_ex} to one or two 
industries, based on their representative occupations: farming and agriculture 
for Topic 2, oil and gas for Topic 3, gambling and hospitality for Topic 4, 
and transportation for Topic 5.
Using the common terms among the representative occupations, we named the 
fifteen industrial topics, which qualitatively match with one of the most 
widely used industrial classification systems — The North American Industry 
Classification System (NAICS).

\begin{table}[!t]
\begin{adjustwidth}{-1.25in}{0in} 
\centering
\small
\caption{
{\bf Industrial Topics and Matching NAICS Code}}
\begin{tabular}{|c|c|c|}
\hline
{\bf Topic} & {\bf Topic Name} & {\bf Corresponding 2-digit NAICS Code} \\ 
&  & {\bf (\% of Traded Industries \cite{delgado2014categorization})}\\ \thickhline
    1 &	Textile & 31(98\%) \\\hline
    2 &	Farming, Agriculture & 11(100\%) \\\hline
    3 &	Oil and Gas & 21(100\%) \\\hline
    4 &	Gambling & 71(88\%) \\\hline
    5 &	Shipping industry & 48(77\%) \\\hline
    6 &	Software / Health and Medical Science / Insurance & 51(78\%), 52(89\%), 54(74\%), 61(65\%)  \\\hline
    7 &	Livestocks / Agriculture, Animal, Life science & 11(100\%), 31(98\%), 54(74\%), 61(65\%) \\\hline
    8 &	Correctional Officers and Specialists & 92(N/A)\\\hline
    9 &	Livestocks, Meat Processing & 11(100\%), 31(98\%) \\\hline
    10 & Furniture, Textile & 31(98\%), 33(98\%) \\\hline
    11 & Chemical & 32(98\%) \\\hline
    12 & Electrical, Computer, Aerospace Engineering & 54(74\%), 33(98\%)\\\hline
    13 & Nuclear Energy / Agriculture & 22(70\%), 11(100\%) \\\hline
    14 & Metal and Plastic & 33(98\%), 32(98\%) \\\hline
    15 & Airline / Hospitality / Urban Services & 48(77\%) \\\hline
\end{tabular}
\begin{flushleft} The name of each industrial topic is decided by analyzing overrepresented 
occupations in the topic. Similarly, the 2-digit NAICS codes are decided by matching 
the corresponding industry categories of overrepresented occupations. 
\end{flushleft}
\label{tab:topic_names}
\end{adjustwidth}
\end{table}

The correspondence between our industrial topics and the industries in NAICS 
reveals the essential features of our approach (Table~\ref{tab:topic_names}). 
Our industrial topics cover most of the occupations in the traded industries, 
which is defined as ``the industries of 2-digit NAICS code where more than 50\% 
of its sub-industries (6-digit NAICS code) are categorized as the trade 
industry \cite{delgado2014categorization}'' (See more detail in Materials and 
methods). All corresponding industries in the existing system are the traded 
industries in which most of its businesses densely locate in a small number 
of specific areas, such as Agriculture, Forestry, Fishing and Hunting (NAICS 
code: 11) and Manufacturing (NAICS code: 31--33). 

Nevertheless, not all industrial topics are clearly matched with one or two 
existing industry categories, and this discrepancy leads us to understand 
the concept of industrial category, not in the product-oriented but, in a
labor-oriented way. First, since each industrial topic is described as the 
weight distribution of all occupations, it is unavailable to match an 
industrial topic to an existing industry category clearly. For example, even the 
industrial topics matched to one 2-digit NAICS code in 
Table~\ref{tab:topic_names} --- such as Topic 1, 2, 3, 4, 5, 8, 11, and 15, 
still those topics include, relatively low but, non-zero weights of the 
occupations related to other industries. It means that our industrial topics 
provide us with additional information about the association between the 
industrial topics, while still allowing us to characterize the core industry.

Also, the existing industry categories in the same industrial topic show the 
relationships between the existing industry categories. For instance, the 
strong relationships between (1) airline and hospitality industries (Topic 15), 
and (2) software, medical science, and insurance industries (Topic 6) are 
well-aligned with the recent study about a landscape of geo-industrial clusters 
in the global economy based on the human capital flow \cite{park2019global}. 
Recent scholarly works about geo-industrial clusters have shown the wide range 
of effects and advantages of having the clusters in the economic growth or 
innovation in a region~\cite{krugman1991, markusen1996, quigley1998, porter2003, 
sorenson2000, ellison2010, hidalgo2007}. The concept of industrial topics 
inferred by the occupational allocation pattern can be better applied to 
future studies about geo-industrial clusters, than the existing industrial 
classification system.

\subsection*{Characterization of urban economy}
Then, how does each industrial topic distributed across the country? Can the 
industrial topics help us understand the structural pattern in the regional 
economy? The application of the topic modeling algorithm on the occupational 
records provides not only the occupational group of an industrial topic 
but also the topical composition of a region (See more detail in Method and 
materials). This topical composition of a region, described as the weights of 
all industrial topics, allows us to answer these questions and 
characterize a regional economy in the labor-oriented perspective.   

\begin{figure}[!t]
\begin{adjustwidth}{-2.25in}{0in}
\centering
\includegraphics[width=1.4\textwidth]{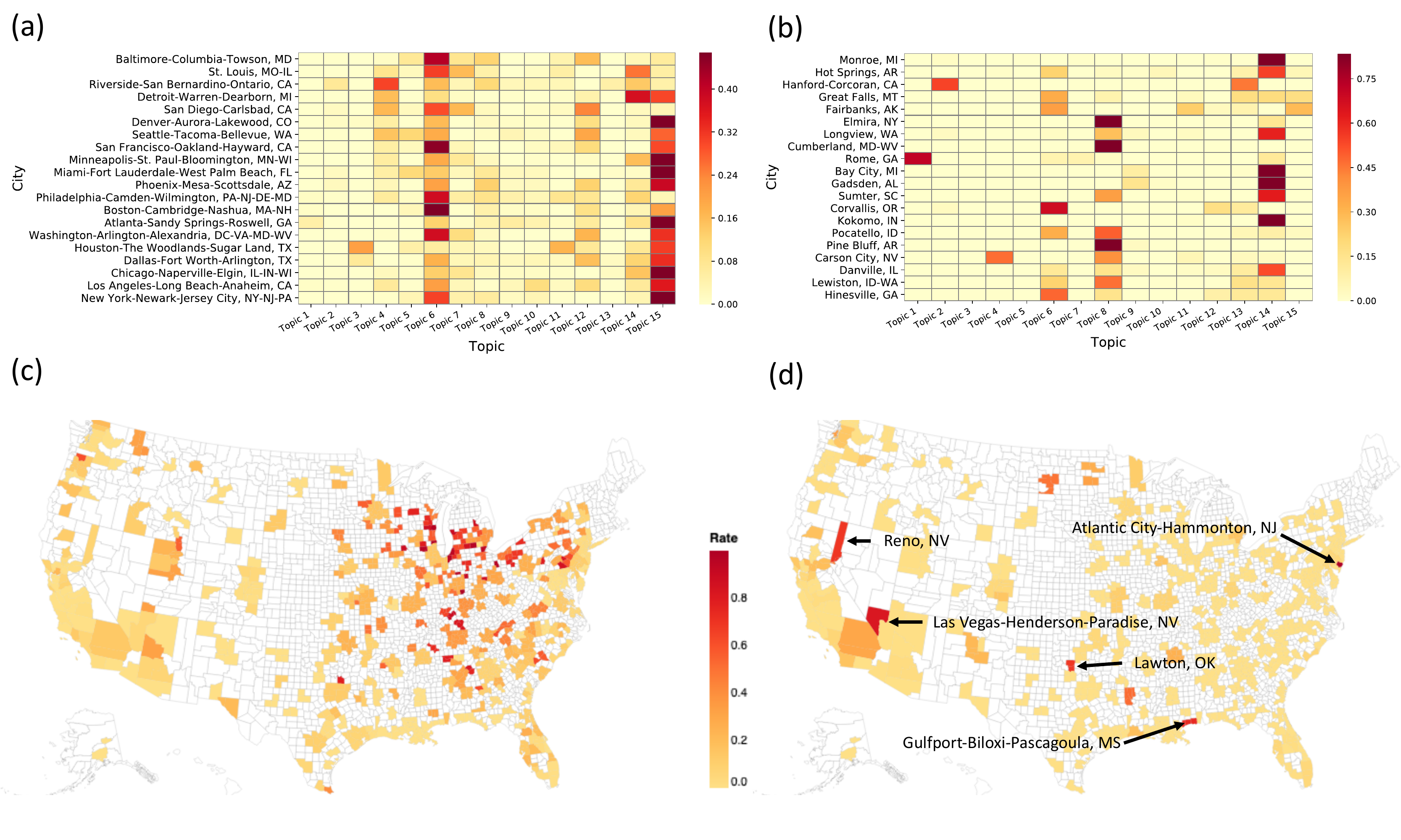}
\caption{{\bf Distribution of industrial topics}
{\bf (a) and (b)} Prevalent 
industrial topics in the ten largest cities (a) and ten smallest cities 
(b) in the U.S. {\bf (c) and (d)} Prevalent regions for 
the industrial topics of metal \& plastic manufacturing (c) and gambling (d). 
For visualization, we match the metropolitan areas in our dataset 
to their corresponding counties, using ``CBSA to FIPS County Crosswalk'' 
generated by the National Bureau of Economic Research.
The maps for all other topics are presented in Supporting Information.}
\label{fig:topic_dist}
\end{adjustwidth}
\end{figure}

In terms of their employment size, we compare the distribution of industrial 
topics in the 20 largest and smallest metropolitan areas. As presented in 
Fig.~\ref{fig:topic_dist}(a) and (b), the comparison between the largest and 
the smallest regional economies shows a clear difference in the economic 
structure depending on the size of a region.  
Large cities commonly have a strong prevalence on Topic 6 (Software / Health 
and Medical Science / Insurance) and 15 (Airline / Hospitality / Urban Services).
However, the balance between the two industrial topics is different for 
different areas, depending on their economic structure. For example, while 
San Diego, Boston, and Baltimore areas have a stronger prevalence of the 
hi-tech topic (Topic 6), Atlanta, Chicago, and Miami areas have a stronger 
prevalence of the hospitality topic (Topic 15). Also, there exist the big 
cities have a balanced prevalence in both topics, such as New York, Washington 
DC, and San Francisco area. On the other hand, small cities have more variation 
in their industrial specialization. Although we can observe a little tendency 
towards correctional institutions (Topic 8) and metal and plastic manufacturing 
(Topic 14), still many of the small cities their specialized industrial 
topics, including textile (Topic 1 for Rome, GA), agriculture (Topic 2 for 
Hanford-Corcoran, CA), and even hi-tech industries (Topic 6 for Corvallis, OR).

\begin{figure}[!ht]
\begin{adjustwidth}{-2.25in}{0in}
\centering
\includegraphics[width=1.4\textwidth]{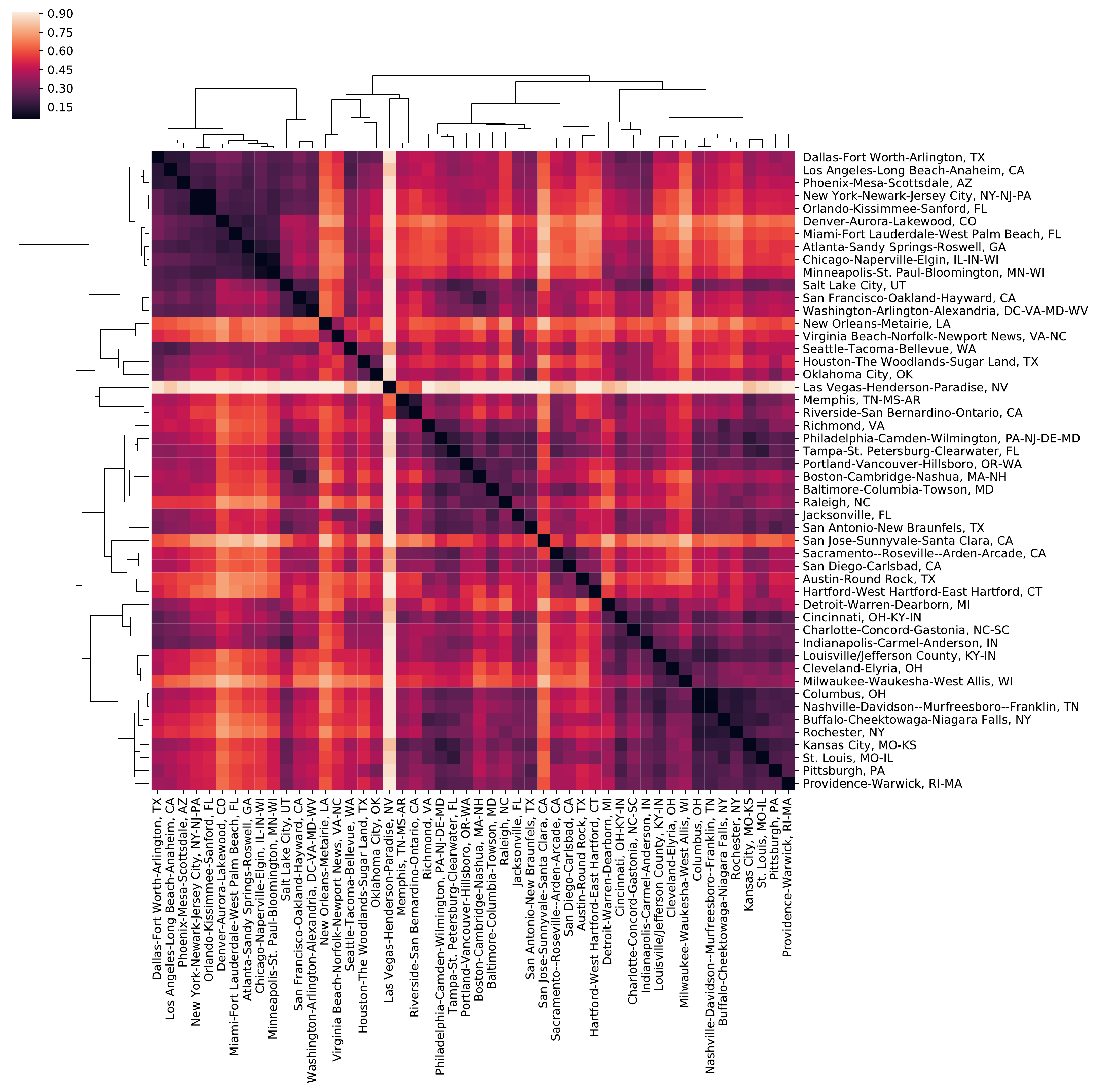}
\caption{Hierarchical clustering of top 50 cities
based on industrial topic composition}
\label{fig:city_clustering}
\end{adjustwidth}
\end{figure}

The geographical distribution of industrial topics provides us with both
the macro-level pattern of an industrial topic and the micro-level information
about the specialized areas across the country.
On the macro-level, some industrial topics densely concentrate in the regions 
that are close to each other, while other industrial topics concentrate in 
multiple regions that are distant from each other across the country.
For example, the industrial topic for metal / plastic
manufacturing (Topic 14, presented in Fig.~\ref{fig:topic_dist}(c)) 
is prevalent in a very specific region of the country, 
that is famous for the manufacturing industry 
in previous studies \cite{helper2012locating, bond2013geographic}. 
This correspondence means that the distribution of the occupations that are either 
directly or indirectly related to metal and plastic manufacturing  
is still well-fitted to the concept of the economic cluster 
\cite{porter1998, porter2003}. 
A few topics, including Topic 2, Topic 15, and Topic 13, have surprisingly large spatial autocorrelation (as measured by the Moran's I statistic. See Append Fig.1 and Table 1) when compared to the spatial autocorrelation observed for the location quotient of employment by NAICS code.

By contrast, the occupational distributions of the other industrial topics
are similar to those of local industries, but on a bigger scale. 
In a city-level, the industrial topic for gambling (Topic 4), 
for example, is considered as 
a traded industry (Table~\ref{tab:topic_names}), since it does not exist 
in all cities 
but concentrates in a small number of cities. However, at the national level,
the industrial topic for gambling is sparsely spread in different parts of
the country, as shown in Fig.~\ref{fig:topic_dist}(d).
Although the levels of concentration are different, each part of the country 
has its representative area for gambling --- such as Atlantic City in East, 
Las Vegas in West, Gulfport in South.
The balanced distribution of the industrial topics that
have been considered as traded industries proposes a critical message: 
the definition of local and traded industry is relative depending on the
scale of our focus. We will talk about this implication more in Discussion. 

Our topic-based representation of the urban economy also 
allows us to cluster the urban areas effectively. By clustering urban areas
based on the similarity in their topical composition, we can categorize the 
urban economies depending on their labor structure. 
Here, as an example, we visualize the hierarchical clustering 
structure of the 50 largest urban areas in Fig.~\ref{fig:city_clustering}
(For the visualization of all cities in our dataset, 
see Appendix A).
In particular, we calculate the cosine distance between the city vectors, 
which are the vectors consists of the weights of the 15 industrial topics,
and cluster the 50 largest cities based on the distance value. 
The heatmap matrix based on the similarity of topical composition between
large US cities provides the insights to understand urban economies
and their groupings. 

First, some cities very unique topical compositions. 
In particular, the topical composition of Las Vegas is unique, so 
distant from those of all other big cities. Considering Las Vegas is the 
only city in the 50 largest US cities among the cities specialized in 
the gambling topic (Fig.~\ref{fig:topic_dist}(d)), we can argue that 
the industrial topic of gambling (Topic 4) is more likely to be prevalent
in the middle- or small-size cities rather than big cities.  
Not as much as Las Vegas, but some other cities, such as 
San Jose-Sunnyvale-Santa Clara, CA, New Orleans-Metairie, LA, and 
Virginia Beach-Norfolk-Newport News, VA-NC, also have the topical composition 
that is quite different from other big cities. 

Second, understanding and comparing the urban economies based on 
topical composition can be utilized for future economic growth.
From the perspective of policymakers or planners of a city, 
the hierarchical clustering provides the information about how likely the 
human capital resource of a city to become those of other cities that
experience economic growth.
For instance, if you are a mayor of Rochester, NY, then the comparison
of topical compositions can provide the information that, for Rochester,
it is easier to encourage the innovation in the industrial
topics that are prevalent in Raleigh, NC than those prevalent in 
San Jose-Sunnyvale-Santa Clara, CA.  
Since the industrial topics are extracted from the occupational distribution,
the ``easier'' means ``more plausible to develop or convert the current human 
capital structure'', through providing job training programs or encouraging 
the migration of specialized human capital.   

\subsection*{Dynamics in industrial topics}

\begin{figure}[!t]
\begin{adjustwidth}{-2.25in}{0in}
\centering
\includegraphics[width=1.4\textwidth]{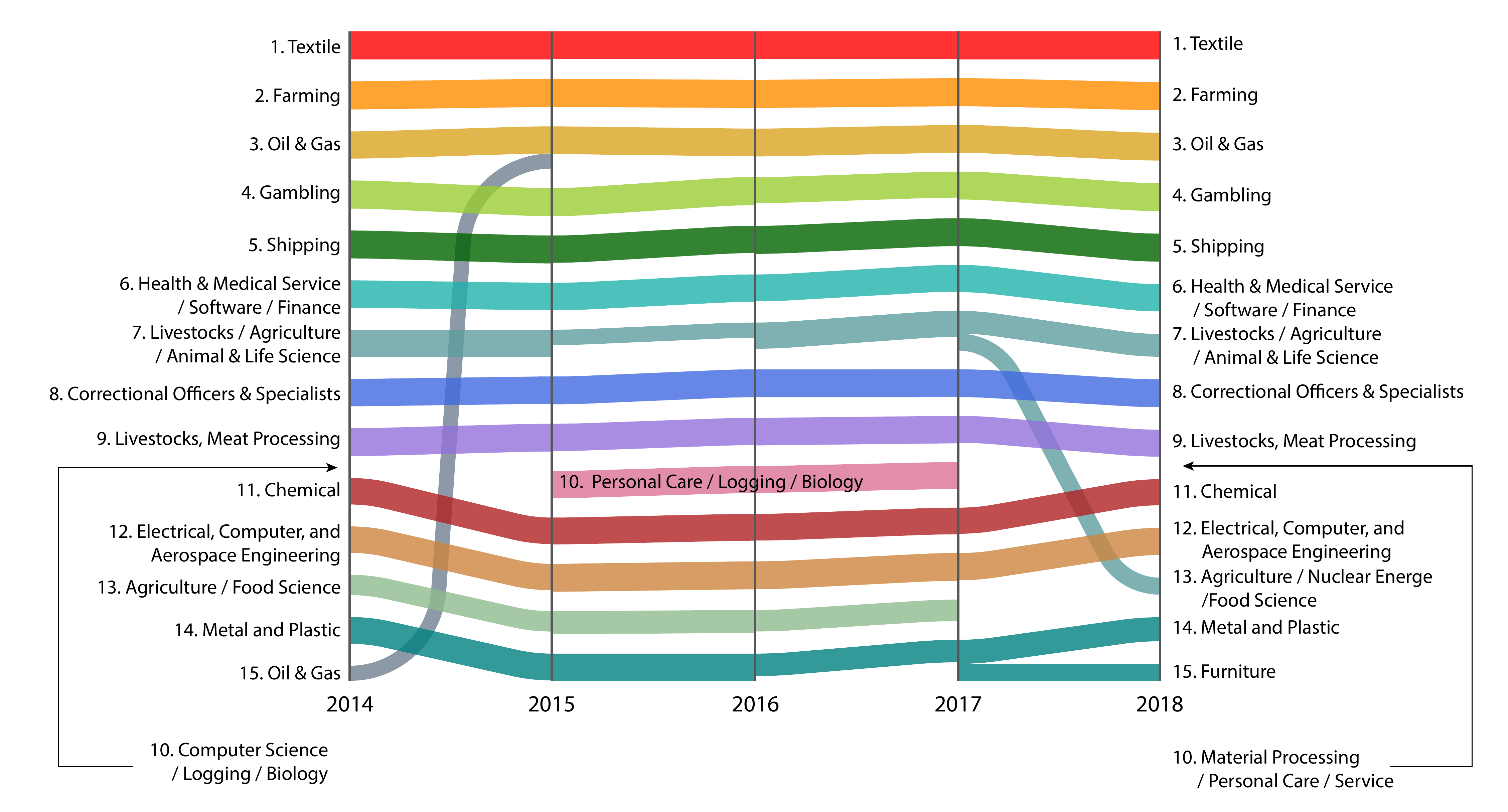}
\caption{{\bf Industrial topic dynamics from 2014 to 2018} The topics are linked 
when the cosine similarity of their occupational distribution vectors is higher 
than 0.5, and the name of the topics are manually assigned based on their 
representative occupations.}
\label{fig:topic_parallel}
\end{adjustwidth}
\end{figure}

Due to its efficiency in the extraction of industrial topics from a corpus of 
regional occupation distributions, the industrial topics also can be used to 
trace the structural dynamics of regional economies across the country. 
How does the industrial structure change over time? How consistent is the 
composition of the industrial topic? Can it also detect a systematic change in the 
industrial structure? In this section, we explore the potential of industrial 
topics to detect the change in the industrial structure by comparing the 
industrial topics in different years.

For the comparison, we first extract the industrial topics from each year's 
OES dataset from 2014 to 2018. 
Then, we align the topics for different years, so that the similar topics 
in different years have the same topic number. We do so by measuring the 
cosine similarity of the occupational distribution vectors between the topics 
of different years (See Materials and methods for detailed information).
The range of cosine similarity is from -1 to 1 --- -1 when two 
vectors opposed, 0 when two vectors oriented at 90° relative to 
each other, and 1 when two vectors with the same orientation --- although we 
here mostly focus on the positive score. In particular, we set a threshold for 
the similarity, $\alpha$, to trace the dynamics of topics over time. For example, 
when $\alpha = 0.5$, we assume that a pair of topics from consecutive years is 
the same or significantly similar topics if the cosine similarity of their 
occupational distribution vectors is higher than 0.5.

The majority of industrial topics are stable.
Fig.~\ref{fig:topic_parallel} visualize the change of industrial topics from 
2014 to 2018 when $\alpha = 0.5$. The thickness of the link between the 
industrial topics of consecutive years represents the similarity of 
occupational distribution. 
Most industrial topics are consistent and linked to similar 
topics of the next year. Except three industrial topics,
12 of 15 industrial topics did not experience a significant change 
in their occupational distribution from 2014 to 2018.

Meanwhile, some industrial topics are merged and separated, depending 
on the economic situation of the occupations that constitute the industrial topics. 
For example, the two industrial topics related to oil and gas in 2014 
(Topic 3 and 15) join together into a single industrial topic in 2015 (Topic 3), 
as the employment in oil and natural gas extraction and support activities in 
the United States declines \cite{mcmanmon2015oil}. By contrast, the industrial 
topic of the occupations related to the nuclear energy industry (Topic 13) as 
representative occupations emerges in 2018, which is well-aligned with the 
time when U.S. nuclear electricity generation surpassed its previous peak 
\cite{scott2019nuclear}. Similarly, the new industrial topic in 2018 related 
to the furniture industry (Topic 15), which concentrates in the North Carolina area 
according to our regional analysis, can be explained by the booming of 
the textile and furniture industry in this area \cite{weisman2017made, 
welton2018north, elliott2019nc}.

\section*{Conclusion}
Here, we proposed a bottom-up approach to identify the industrial topics 
that allows us to detect the specialization of the regional economy 
from occupation distribution records. 
In particular, we applied a topic modeling algorithm to extract 
the latent allocation patterns of the occupational distribution. 
The proposed model has several advantages over conventional analysis: 
(1) the model is flexible with new data (i.e., an updated occupation employment 
data set), and (2) the model efficiently detects structural change of 
the economy. Our result shows that the groups of occupations, similar to 
the industry mix, also demonstrates a substantial 
geographic concentration. 

As the first attempt to apply a topic modeling algorithm to defined and extract 
industrial topics as a group of related occupations, 
our approach has limitations. First, while our industrial topics are
more flexible and adjustable to the contemporary economic structure, still
they are constrained by the less flexible list of occupations. 
Our approach cannot detect the dynamics of industrial structure driven by
new occupations, for example, data scientists or AI specialists. 
This limitation can be overcome by leveraging user-generated real-time-based 
skill or occupation information, such as that from 
job searching and hiring services. 
Second, since we extract industrial topics based on each region's overrepresented
occupations,
our industrial topics underemphasize local industries, whose occupations
evenly appear across regions in general. 
We expect local industries to be extracted using a different 
measurement for occupations from TF-IDF, separately from traded industries. 

Nevertheless, our results agree with the recent clustering theory in regional science and 
urban economics, indicating that the regional economy does show clear clustering 
patterns. It helps us identify how the key industry sectors of cities are 
performing and how industrial signatures differ from city to city 
\cite{feldman1999}. The extracted topics also show a clear geographical 
concentration. By tracing the change of each topic, we found that 
the weight of topics shows a great variation over the last four years. 
This change also corresponds to the development of cities. The topic 
distribution offers us a new feature to study a region's economic performance. 
Our study indicates that the occupation components better 
reveal the structure of the regional economy. We hope our work could innovate more 
and new approaches and measures to define regional economy \cite{guzman2015} 
and help research in city science \cite{bettencourt2007,batty2008}.

\section*{Materials and methods}
\subsection*{Dataset}
We analyze the occupational distribution using the Occupational Employment 
Statistics (OES) Survey data, which is a semi-annual survey conducted by Bureau 
of Labor Statistics to estimate employment and wage~\cite{OES_data}. 
The dataset provides the number of employment and average wage for each occupation 
at the spatial level of Metropolitan Statistical Area (MSA).

The Occupational Employment Statistics (OES) data, collected by Bureau of Labor 
Statistics through a series of mail surveys, measure occupational employment 
and rates of wage and salary in non-farm establishments. The dataset also does 
not include self-employed persons. The OES data is available at different levels: 
nation as a whole, state, metropolitan/non-metropolitan area, etc. 
We refer readers to \url{http://www.bls.gov/oes/oes_emp.htm} for further 
details of the data set. In this study, we use the occupational employment and 
annual wage estimates at the level of MSA. In this study, we used the datasets 
during the five most recent years --- from 2014 to 2018 --- at the time of our 
analyses.
As a result, our 5-year dataset includes the distribution of 829 occupations 
across 361 MSAs in the US from 2014 to 2018.
Since there exist the MSAs newly appeared and disappeared during our target 
periods. For the consistency issue, we only include the MSAs whose major 
corresponding regions are consistently appeared throughout the period,
while manually matching the old and new area codes. For instance, 
we match the area code ``31084 - Los Angeles-Long Beach-Glendale, CA Metropolitan Division''
in 2017 with ``31080 - Los Angeles-Long Beach-Anaheim, CA'' in 2018. 
The complete matching table is presented in Supporting information.

\subsection*{Term frequency–inverse document frequency (TF-IDF)}
\label{method:tf_idf}

We use TF-IDF, ``term frequency–inverse document frequency'', 
to score the importance of an occupation in a region based on how many 
employments the occupation has in that region and across all the regions. 
The intuition for this measure is: If an occupation appears frequently in a 
region, then it should be important and we should give that occupation a high 
score. But if the occupation appears in too many other regions, it’s probably 
not a unique identifier, therefore we should assign a lower score to that 
occupation. The math formula for this measure:

\begin{equation}
\begin{split}
    & tf_{o,r} = f_{o,r} \\
    & idf_{o, R} = log\frac{N}{|{r \in R:o \in r}|}, and \\
    & tf-idf_{o,r} = tf_{o,r} \cdot idf_{o, R}  
\end{split}
\end{equation}

where $o$, $r$, $R$, and $N$ represent an occupation, a region, all regions, and 
the total number of regions, respectively.

\subsection*{Non-negative Matrix Factorization (NMF)}
Nonnegative Matrix Factorization (NMF) is one of the widely used topic 
modeling algorithms. 
NMF, also called \textit{positive matrix factorization} 
\cite{paatero1994positive} or \textit{nonnegative matrix approximation} 
\cite{sra2006generalized}, is a computational technique to reduce dimensionality 
to analyze a high-dimensional data as a part based \cite{lee2001algorithms}. 
Along with other diverse applications, including those in astrophysics 
\cite{ren2018non}, bioinformatics \cite{murrell2011non}, image processing 
\cite{lee1999learning}, and recommendation system \cite{gemulla2011large}, 
NMF has been widely applied to topic extraction based on the allocation of 
words across a corpus of documents \cite{lee1999learning, bao2014topicmf}. 

The goal of NMF is to find a non-negative matrix which can approximately 
reproduce a given matrix, while having a lower dimension.
Given a matrix $X$ with nonnegative elements, NMF algorithm 
aims to finds a decomposition of $X$
into two matrices --- $W$ and $H$ having non-negative elements, such that

\begin{equation}
X \approx W H
\end{equation}

\textit{W} and \textit{H} can be found by optimizing by 
the squared Frobenius norm, which is an extension of the 
Euclidean norm for matrices to calculate the distance, such that:

\begin{equation}
    d_{Fro}(X,WH) = \frac{1}{2}||X-WH||^{2}_{Fro} = \frac{1}{2}\Sigma_{i,j}(X_{ij}-WH_{i,j})^{2}
\end{equation}

Similar to other dimension reduction methods, such as 
Principal Component Analysis (PCA) or Latent Dirichlet allocation (LDA), NMF 
requires to set the number of topics. 
Since our study focus on the traded industry, we set the latent number of 
topics as the number two-digit NAICS codes that are considered as traded 
industries. Delgado et al. recently calculate the percentage of sub-industries 
(6-digit NAICS code) classified as the traded industries, for each 2-digit 
NAICS code \cite{delgado2014categorization}. According to their calculation, 
15 of 23 2-digit NAICS codes have more than 50\% of their sub-industries that 
are traded industries. Hence, we set 15 as the latent number of topic here, 
although the number can be changed for future studies. 

\subsection*{Alignment of topics across years}
For the static analyses, we extract the 
industrial topics from the integrated dataset from 2014 to 2018.
To trace the dynamics of the industrial topics during the five years, 
we extract the industrial topics separately using the dataset of each year.

The NMF algorithm does not guarantee that the order of the topics are 
kept for the similar datasets. That means, for example, the topic strongly 
related to the textile industry is marked as the first topic in the result for 
2014 dataset, while the topic is marked as the third topic in the result for 
2015. Hence, we align the order of the topics between years, by calculating 
the similarity between the topics in the consecutive years. In particular, we 
use the cosine distance between the occupational distribution vectors of the 
two topics in different years as the measurement of the similarity.  

\section*{Acknowledgments}
We thank Yong-Yeol Ahn for helpful comments and discussion. This work was supported by the National Research Foundation of Korea Grant funded by the Korean Government (NRF-2018S1A3A2075175), and London Mathematical Laboratory. 

\bibliography{main_arxiv}
\section*{Supporting information}
\begin{table}[!h]
\centering
\begin{tabular}{|c|c|c|c|c|}
\hline
{\bf Area Code} & {\bf Area Code} & {\bf Area Code} & {\bf Area Code} & {\bf Area Code}\\ 
{\bf in 2014} & {\bf in 2015} & {\bf in 2016} & {\bf in 2017} & {\bf in 2018}\\ \thickhline
16980 & 16980 & 16974 & 16974 & 16980\\\hline
35620 & 35620 & 35614 & 35614 & 35620\\\hline
31100 & 31080 & 31084 & 31084 & 31080\\\hline
41860 & 41860 & 41884 & 41884 & 41860\\\hline
47900 & 47900 & 47894 & 47894 & 47900\\\hline
33100 & 33100 & 33124 & 33124 & 33100\\\hline
14060 & 14010 & 14010 & 14010 & 14010\\\hline
19100 & 19100 & 19124 & 19124 & 19100\\\hline
19820 & 19820 & 19804 & 19804 & 19820\\\hline
29140 & 29200 & 29200 & 29200 & 29200\\\hline
37980 & 37980 & 37964 & 37964 & 37980\\\hline
42060 & 42200 & 42200 & 42200 & 42200\\\hline
42660 & 42660 & 42644 & 42644 & 42660\\\hline
71650 & 71650 & 71654 & 71654 & 71650\\\hline
\end{tabular}
\caption{
{\bf Matching table of area codes} For the areas 
whose codes and corresponding areas were slightly changed during the period, 
we manually match the area codes if their major corresponding areas are same.
}
\end{table}

\begin{figure}[!ht]
\begin{adjustwidth}{-2.25in}{0in}
    \centering
    \includegraphics[width=1.3\textwidth]{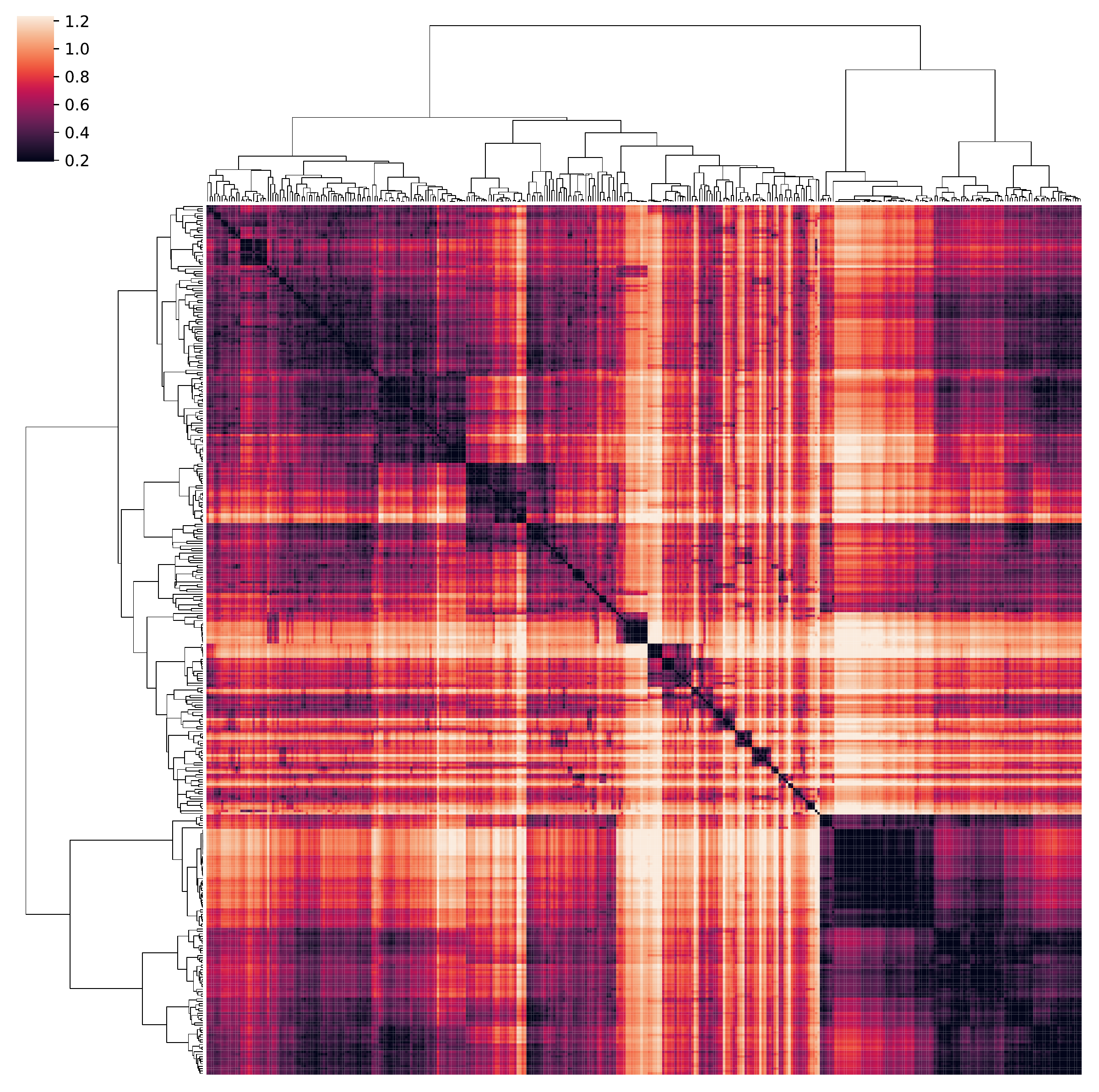}
\caption{Hierarchical clustering of top 50 cities based on industrial topic composition}
\label{fig:city_clustering}
\end{adjustwidth}
\end{figure}

\begin{figure}[!ht]
    \centering
    \includegraphics[width=\textwidth]{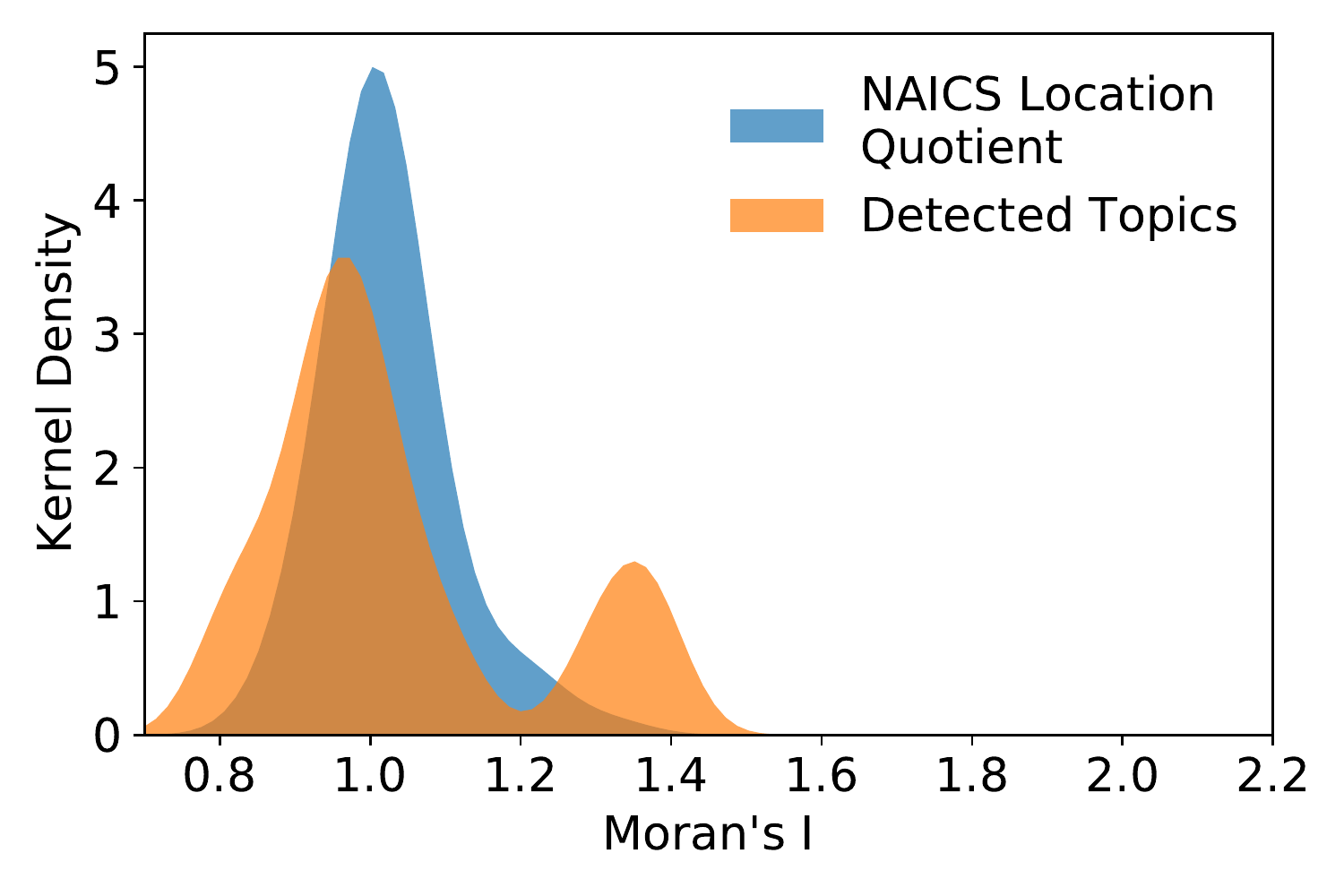}
\label{fig:spatialAutocorrelationDensity}
\caption{
    The distribution of the Moran's I test statistic for spatial autocorrelation for the location quotient of employment by NAICS sector (blue) and the topics detected by our model (orange).
    A few of our detected topics exhibit greater spatial autocorrelation than would be expected based on the spatial autocorrelation of NAICS sector employment.
}
\end{figure}

\begin{table}[h]
    \centering
    \begin{tabular}{lllr}
Rank &     Topic &                                              Topic Label &  Moran's I \\ \hline
1  &   topic\_1 &                                            Textile &   0.808272 \\
2  &  topic\_10 &                                 Furniture, Textile &   0.839089 \\
3  &  topic\_14 &                                  Metal and Plastic &   0.903845 \\
4  &   topic\_8 &              Correctional Officers and Specialists &   0.926108 \\
5  &   topic\_7 &     Livestocks / Agriculture, Animal, Life science &   0.931856 \\
6  &  topic\_11 &                                           Chemical &   0.963046 \\
7  &   topic\_5 &                                  Shipping industry &   0.964980 \\
8  &  topic\_12 &        Electrical, Computer, Aerospace Engineering &   0.976718 \\
9  &   topic\_9 &                        Livestocks, Meat Processing &   0.997920 \\
10  &   topic\_6 &  Software / Health and Medical Science / Insurance &   1.014528 \\
11 &   topic\_3 &                                        Oil and Gas &   1.044966 \\
12 &   topic\_4 &                                           Gambling &   1.106953 \\
13 &  topic\_13 &                       Nuclear Energy / Agriculture &   1.299990 \\
14 &  topic\_15 &             Airline / Hospitality / Urban Services &   1.356454 \\
15 &   topic\_2 &                               Farming, Agriculture &   1.379097 \\
\end{tabular}
\caption{
    Detected topics ranked by spatial autocorrelation.
    Employment in Topic 2 (Farming, Agriculture), Topic 15 (Airline/Hospitality/Urban Services), and Topic 13 (Nuclear Energy/Agriculture) have especially large spatial autocorrelation compared to the expected spatial autocorrelation of the location quotient of employment for NAICS sectors.
}
\end{table}

\end{document}